\documentclass[aps,prb,twocolumn,showpacs,superscriptaddress,10p,longbibliography]{revtex4-1}
\usepackage{amsmath}
\usepackage{amssymb}
\usepackage[colorlinks=true,linkcolor=red, citecolor = magenta]{hyperref}
\usepackage{epsfig}
\usepackage{bm}
\usepackage[up]{subfigure}
\usepackage{bbm}
\usepackage{ulem}
\usepackage{graphicx}
\usepackage{subfigure}
\usepackage{color}
\usepackage{braket}
\usepackage{dsfont}
\usepackage{footnote}
\usepackage{textcase}
\usepackage[shortlabels]{enumitem}
\usepackage{float}
\usepackage{mathtools}
\begin{document}
\title{Spectral Magnetization Ratchets with Discrete Time Quantum Walks}

\author{A. Mallick}
\affiliation{Center for Theoretical Physics of Complex Systems, Institute for Basic Science (IBS), Daejeon 34126, Republic of Korea}

\author{M. V. Fistul}
\affiliation{Center for Theoretical Physics of Complex Systems, Institute for Basic Science (IBS), Daejeon 34126, Republic of Korea} 
\affiliation{National University of Science and Technology ``MISIS", Russian Quantum Center,  Moscow 119049, Russia}

\author{P.  Kaczynska}
\affiliation{Center for Theoretical Physics of Complex Systems, Institute for Basic Science (IBS), Daejeon 34126, Republic of Korea}
\affiliation{Faculty of Physics, University of Warsaw, 
 Warsaw 02-093, Poland}

\author{S. Flach}
\affiliation{Center for Theoretical Physics of Complex Systems, Institute for Basic Science (IBS), Daejeon 34126, Republic of Korea}

\date{\today}

\begin{abstract}
 We predict and theoretically study in detail the ratchet effect for the spectral magnetization of {\it periodic} discrete time quantum walks (DTQWs) ---
a repetition of a sequence of $m$ different DTQWs.
These generalized DTQWs are achieved by varying the corresponding coin operator parameters periodically with discrete time. 
We consider periods $m=1,2,3$.
The dynamics of $m$-periodic DTQWs is characterized by a two-band dispersion relation $\omega^{(m)}_{\pm}(k)$, where $k$ is the wave vector.
We identify a generalized parity symmetry of $m$-periodic DTQWs. The symmetry can be broken for $m=2,3$ by proper choices of the coin operator
parameters.
 The obtained symmetry breaking results in a ratchet effect, i.e. the appearance of a nonzero spectral magnetization $M_s(\omega)$. 
 This ratchet effect can be observed in the framework of continuous quantum measurements of the time-dependent correlation function of periodic DTQWs.
\end{abstract}

%================================================
\maketitle

\section{Introduction}
Transport properties of particles and waves in spatially periodic structures that are driven by external time-dependent forces manifestly depend on the space–time symmetries 
of the underlying equations of motion. A systematic analysis of these symmetries uncovers the conditions necessary for their violation and the appearance of 
the ratchet phenomenon to e.g. explain rectification of currents \cite{reimann2002brownian,julicher1997modeling,hanggi2005brownian,denisov2014tunable}.
Such phenomena have been predicted and studied in detail in various Hamiltonian and dissipative systems, for single particle \cite{flach2000directed} and for many-body interacting systems 
\cite{flach2002broken}.  
Ratchets have been observed in various solid-state \cite{carapella2001ratchet},
optical \cite{zhang2015experimental}, chemical and biological \cite{julicher1997modeling} systems. 

Classical ratchet experimental platforms are modeled with a set 
of coupled nonlinear differential equations whose parameters vary in space and time.
The ratchet effect results from broken spatio-temporal symmetries of the differential equations.
Spatio-temporal symmetries typically involve discrete shift and parity operations \cite{denisov2014tunable}.

The quantum ratchet concept was predicted theoretically \cite{reimann1997quantum,denisov2007quantum,denisov2007periodically} 
and successfully implemented for a variety of different  quantum systems platforms \cite{majer2003quantum,salger2009directed,drexler2013magnetic}.
Quantum ratchets are typically described by quantum Hamiltonian systems which are periodically driven in time.
The main body of studies was devoted to rectifying charge currents. An incoherent ratchet effect for driven and damped spins was reported in
Refs.~\cite{SFAAO_2001, SFAEMAAO_2002}. To the best of our knowledge, spectral magnetization ratchets,
i.e. frequency-selective magnetization ratchets for coherent non-dissipative quantum spin systems, were not considered so far.

To address the coherent quantum spin ratchet dynamics, we use the platform of discrete time quantum walks (DTQW)\cite{PhysRevA.48.1687}. 
The DTQW is a spatio-temporal unitary map developed for quantum computing \cite{Lovett2010, Singh2019}, which is obtained from a repeating  sequence of 
coin and shift operators acting on a two-level (spin $1/2$) system network. Recently such platforms turned into a
playground to study various interesting physical phenomena, e.g. single particle and many body Anderson localization 
\cite{vakulchyk2017anderson, crespi2013anderson, vakulchyk2019wave, crespi2013anderson},
topological phenomena \cite{kitagawa2010exploring, asboth2012symmetries, PhysRevA.92.052311},
propagating solitons \cite{Maeda2019}, relativistic Dirac particle systems \cite{Arindam_CM_2016, Mallick_2019, Arnault_Perez_2019} etc.
DTQWs have been experimentally implemented using ion-traps, photonic crystals, NMR \cite{du2003experimental_NMR}, cavity-QED \cite{di2004cavity} etc. 
DTQW ratchets \cite{Meyer_parrondo_2003, 2017AnP...52900346C, Jishnu_2018} were introduced for directed currents.

In order to obtain the spectral magnetization effect with DTQWs we introduce their generalization---$m$-periodic DTQWs ---
a repetition of a sequence of $m$ different DTQWs.
These generalized DTQWs are achieved by varying the corresponding coin operator parameters periodically as functions of the discrete time. 
We identify various symmetries of $m$-periodic DTQWs, and outline ways to break them for $m=2,3$. 
 This ratchet effect can be observed in the framework of continuous quantum measurements of the time-dependent correlation function of periodic DTQWs.

The paper is organized as follows. We first
introduce the model and dynamic equations for $m$-periodic DTQWs.  We proceed with defining 
dispersion relations, eigenvectors, and magnetization properties. We continue to define the generalized parity symmetry.
For $m=1,2,3$ we analyze the conditions under which the generalized parity is broken, derive the symmetry breaking conditions, and obtain 
a spectral magnetization ratchet for $m=2,3$.
Finally we discuss observation 
methods and conclude.
\section{$m$-periodic Discrete Time Quantum Walks}
\label{define_DQW}
We consider a single particle $m$-periodic DTQW which is defined on
a lattice of $N$ sites. The quantum-mechanical dynamics of arbitrary DTQW is characterized by 
two-component wave functions $\ket{\psi(n,t)}$ = $\left(\psi_+(n,t),~ \psi_-(n,t)\right)^T$
which depend on both site $n$ and discrete time
$t$.
The discrete time-dependent probability amplitude for the whole system is characterized by the state 
$\ket{\psi(t)}$ = $\sum_{n = 1}^N  \ket{n} \otimes \ket{\psi(n,t)}$.  
The $m$-periodic dynamics of such wave functions is determined by coin operators $\hat C_\ell$, with the temporal index $\ell$ varying from $1$ to $m$, and a shift operator, $\hat S$ acting on the state as follows: 
\begin{figure}[!h]
 \includegraphics[width = 7.2cm]{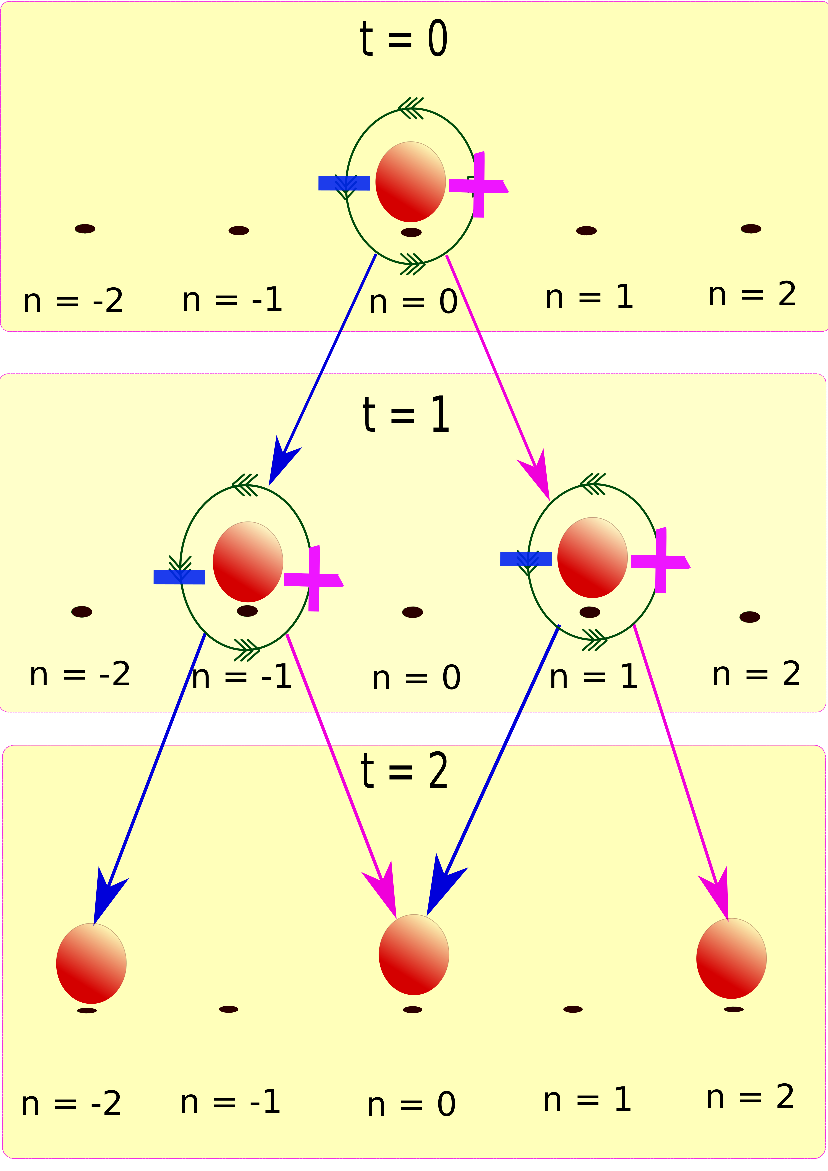}
 \caption{Schematic representation of the 1-periodic DTQW.
 The lattice sites $n$ are denoted by black dots, time flows from top to bottom following the arrows. 
 The DTQW wavefunction $\ket{\psi(n,t)}$ is initialized at $t=0$ and site $n=0$ and evolves as denoted by straight arrows.
 The ``$\pm$'' signs indicate two spin-components (magenta (gray) and blue (light gray)). 
 The double-arrow-decorated circles indicate coin operations which rotate the 
 spin-components at each site. The  straight 
 arrows indicate the shift operations with the ``$+$'' component shifted along the magenta (gray) arrow and the ``$-$'' component along the blue (light gray) arrow.} \label{Schematic}
\end{figure}

\begin{eqnarray}
\label{evolution} 
\ket{\psi(t+\ell)} = \hat S \cdot \hat C_\ell~ \ket{\psi(t+\ell-1)},~~\ell=1,...m~.
\end{eqnarray}
We consider site-independent coin operators $\hat{C}_{\ell}$:
\begin{eqnarray}\label{Coin1}
    \hat{C}_\ell= \mathds{1} \otimes e^{i\varphi_\ell}
    \begin{pmatrix}
        e^{i\varphi_{1,\ell}}\cos\theta_\ell   & e^{i\varphi_{2,\ell}}\sin\theta_\ell \\
        -e^{-i\varphi_{2,\ell}}\sin\theta_\ell & e^{-i\varphi_{1,\ell}}\cos\theta_\ell
    \end{pmatrix},
\end{eqnarray}
where $\mathds{1}$ is the identity operator on position space, i.e. with rank $N$ for a total number of $N$ sites. 
The DTQW dynamics at each time $t$ is determined by four angles: $\varphi, \varphi_1, \varphi_2, \theta$.
These angles can be related to the action of a potential energy, external and internal synthetic magnetic flux, and a kinetic energy, respectively \cite{vakulchyk2017anderson}. As outlined below, the potential energy angle $\varphi$ turns irrelevant, and we will always set it to zero: $\varphi \equiv 0$.
The shift operator couples neighboring sites by transferring the $\psi_+(n,t)$ components one step to the right,
and the $\psi_-(n,t)$ components to the left:
\begin{eqnarray}\label{Shift1}
    \hat{S} = \sum_n \ket{n}\bra{n+1} \otimes \ket{-}\bra{-} \; +\; \ket{n}\bra{n-1} \otimes \ket{+}\bra{+}. \nonumber\\ 
\end{eqnarray}
We then arrive at the generalized evolution operator of 
$m$-periodic DTQWs:
\begin{align}
 \hat{U}^{(m)} = \prod_{l = 1}^m \hat{U}_\ell = \prod_{l = 1}^m \hat S \cdot \hat C_\ell~.
\end{align}

The schematic of the DTQW evolution is presented in Fig.~\ref{Schematic}.

Translational invariance of the evolution operator 
$\hat U_\ell$ = $\hat S \cdot \hat C_\ell$ allows to apply Bloch's theorem and
to expand the wave function in the plane wave basis as $\ket{\psi(n,t)}=\frac{1}{\sqrt{N}}\sum_k e^{ikn} \ket{\psi(k,t)}$ 
where $\ket{\psi(k,t)} = \left(\psi_+(k,t),~ \psi_-(k,t)\right)^T$ is the two-component wave function in momentum space. 
The dynamics of an $m$-periodic DTQW in $k$-space follows as
\begin{eqnarray}\label{DynamicEquation}
    \ket{\psi(k,t+m)} = \prod_{\ell=1}^m \hat{U}_\ell(k) \ket{\psi(k,t)}~.
\end{eqnarray}
The evolution operator for a single $m$-period can be written as
$\hat{U}^{(m)}$ = $\sum_k \ket{k}\bra{k} \otimes \hat{U}^{(m)}(k)$ 
= $\sum_k \ket{k}\bra{k} \otimes \prod_{\ell=1}^m \hat{U}_\ell(k)$ where
\begin{eqnarray}\label{U-operator} 
    \hat{U}_\ell(k)=e^{i\varphi_\ell}
    \begin{pmatrix}
        e^{i\varphi_{1,\ell}-ik}\cos\theta_\ell   & e^{i\varphi_{2,\ell}-ik}\sin\theta_\ell \\
        -e^{-i\varphi_{2,\ell}+ik}\sin\theta_\ell & e^{-i\varphi_{1,\ell}+ik}\cos\theta_\ell
    \end{pmatrix}.~
\end{eqnarray}

\section{Dispersion relations and eigenvectors}\label{define_dispersion}

The solution of Eq.~(\ref{DynamicEquation}) is written as  
$\ket{\psi(k,t)}$ = $e^{-i\omega^{(m)} t} \ket{\psi(k,\omega^{(m)})}$,
where $\ket{\psi(k,\omega^{(m)})}$ = $(\psi_+(k,\omega^{(m)})$, $\psi_-(k,\omega^{(m)}))^T$ is the corresponding two component wave function in momentum and frequency space. The frequency $\omega^{(m)}$ is a function 
of the momentum $k$ and can take two values for fixed value of $k$, i.e.~$\omega^{(m)}_{\pm}(k)$. This follows directly from having two levels (degrees of freedom) per lattice site (unit cell) which dictates a band structure with two bands.
From Eqs.~(\ref{DynamicEquation}) and (\ref{U-operator}) it follows that nonzero values of the angles $\varphi_{\ell}$ 
result in a shift of $\omega^{(m)}_{\pm}$ only. This is similar to a potential which is constant in space and only shifts the energy of a quantum system.Therefore we set 
$\varphi_{\ell}=0$ for all $l$.

The evolution operator 
\begin{eqnarray}\label{Operator}
\hat{U}^{(m)}(k)=
\prod_{\ell=1}^m \hat{U}_\ell(k) =
\begin{pmatrix}
u^{(m)}_{11}~&~ u^{(m)}_{12}\\ 
-\left[u^{(m)}_{12}\right]^\star~&~\left[u^{(m)}_{11}\right]^\star
\end{pmatrix}
\end{eqnarray}
is a unitary matrix, whose eigenvalues 
yield the dispersion relation of $m$-periodic DTQWs:
\begin{eqnarray}\label{GeneralDispRel}
  \omega^{(m)}_{\pm}(k)=\pm \frac{1}{m} \arccos\left(\Re e\left[u^{(m)}_{11}(k)\right]\right).
\end{eqnarray}
For each value of the wave number $k$ we find two frequencies with opposite values. Thus the band structure of any $m$-periodic DTQW is given
by two bands which are symmetry related by (\ref{GeneralDispRel}).
Because of Bloch's theorem, the matrix elements $u^{(m)}_{\alpha \beta}$ for all $\alpha$, $\beta$ 
in Eq.~(\ref{Operator})
are periodic functions of $k$.
Hence $\omega^{(m)}_{\pm}(k)$ is also a periodic function of $k$.
The corresponding eigenvectors are given by $\ket{k} \otimes \ket{\psi(k,\omega^{(m)}_\pm)}$ with the 
%coin
spinor part
\begin{align}\label{Generaleigvec}
 \ket{\psi(k,\omega^{(m)}_\pm)}=  \frac{\left(i u^{(m)}_{12},~~\Im m \left[u^{(m)}_{11}\right] + \sin\left[m\omega^{(m)}_\pm\right] \right)^T}
{\sqrt{\left| u^{(m)}_{12} \right|^2 + \left|\Im m \left[u^{(m)}_{11}\right] + \sin\left[m \omega^{(m)}_\pm\right] \right|^2}}~.
\end{align}

\section{Magnetization}

The magnetization of an eigenstate measures the population imbalance between the upper and lower levels. It is obtained from the expectation value
of the magnetization operator $\hat{M} = \mathds{1} \otimes \hat \sigma_3$ as
\begin{multline}\label{magne_eigen1}
 M_{\pm}(k)=(\braket{k|\mathds{1}|k} \braket{\psi(k, \omega^{(m)}_\pm| \hat \sigma_3 |\psi(k, \omega^{(m)}_\pm)} \\
 \equiv |\psi_+(k, \omega^{(m)}_\pm)|^2-|\psi_-(k, \omega^{(m)}_\pm)|^2   ~. 
\end{multline}
With Eqs.~(\ref{Generaleigvec}) and (\ref{magne_eigen1}) it follows 
\begin{align}\label{Magnetization-Gen-1}
M_{\pm}(k)
= \frac{-\Im m[u^{(m)}_{11}(k)]}{\sin [m\omega^{(m)}_{\pm}(k)]} \equiv \mp \frac{\Im m[u^{(m)}_{11}(k)]}{  \sqrt{1-  (\Re e [u^{(m)}_{11}(k)])^2 } }\;.
\end{align}
Note that $\pm$ refers to the upper respectively lower branch
of the two band dispersion relation.

The above result (\ref{Magnetization-Gen-1}) is quite remarkable and can be used for a number of conclusions. The magnetization of an eigenstate is entirely defined by the matrix element
$u^{(m)}_{11}(k)$ of the evolution operator $\hat{U}^{(m)}(k)$, c.f. (\ref{Operator},\ref{GeneralDispRel}). It follows that 
the upper and lower branch magnetizations are opposite to each other: $M_{+}(k)= - M_{-}(k)$. Exciting a monochromatic (single wavelength) mix of states with one value of $k$ and equal weights of eigenstates yields zero monochromatic magnetization
\begin{equation}
M_{k} = M_{+}(k)+M_{-}(k) = 0
\label{Mk}
\end{equation}
 for any $m$-periodic DTQW. 
Thus also the total magnetization $M_{tot}$ - the 
sum over the magnetization values for all eigenstates (with equal weight) vanishes pairwise for each $k$ and is exactly zero for any $m$-periodic DTQW :
\begin{equation}
M_{tot} =  \sum_k M_k =  0 \;.
\label{mtot}
\end{equation}

At variance to the above, the spectral magnetization $M_s(\omega)$ measures the average magnetization of all eigenstates with $\omega_{\pm}(k_l) = \omega$.
Different eigenstates with identical frequency can be excited using spectroscopic methods, as we will show below.
Due to the fact that $\Re e ~u_{11}(k)$ and consequently $\omega_{\pm}(k)$ are periodic functions of $k$, the spectral magnetization will average over a discrete set of eigenstates counted by the integer $l$:
\begin{equation}
M_s(\omega) = \sum_{k_l} M_{\pm}(k_l) \;,\; \omega = \omega_{\pm}(k_l)\;.
\label{mspec}
\end{equation}
For a fixed value of $\omega$, the denominator in Eq.(\ref{Magnetization-Gen-1}) is invariant for all allowed values of $k_l$. Further, since $\omega_{\pm}(k)$ is a periodic function in $k$, the set $\{k_l \}$ contains an even number of states. 
The rest of this work is devoted to answering the question, under which conditions a generalized parity symmetry will hold such that the set $\{ k_l \}$ will have symmetry-related
pairs of states for which the magnetization vanishes pairwise. The breaking of that generalized parity symmetry will then lead to a nonzero spectral magnetization. We coin this effect {\sl spectral magnetization ratchet}.

\section{Generalized parity symmetry}

The spectral magnetization ratchet 
requires the breaking of the \textit{generalized parity symmetry}.
If the ratchet effect is absent, the $m$-periodic DTQW is invariant under the action of the generalized parity symmetry operation:
\begin{align}\label{ParitySymm}
\hat{U}^{(m)} = (\mathcal{P} \otimes \mathcal{G}) \cdot  \hat{U}^{(m)} \cdot (\mathcal{P} \otimes \mathcal{G})^{\dagger},
\end{align}
where $\mathcal{P}$ is an operator inducing reflection in momentum-space around some wave number $K$,
and $\mathcal{G}$ is an operator inducing spin flips with an additional phase shift $G$: 
\begin{align}
\mathcal{P} &= \sum_k \ket{2 K - k}\bra{k}
 = \sum_x e^{- 2 i K x} \ket{-x}\bra{x} \;, \nonumber\\ 
\mathcal{G} & =\left(\begin{array}{cc}
            0 & -e^{ i G}\\ 1 & 0 \\
           \end{array}\right). \label{PundG}
\end{align}
If existing, the values of $K$ and $G$  will depend on the particular parameters of the $m$-periodic DTQW - hence the term {\sl generalized} parity.
In the presence of that symmetry, for each eigenstate $\ket{k} \otimes \ket{\psi(k, \omega^{(m)}_\pm)}$
there exists another eigenstate $\ket{2K - k} \otimes \ket{\psi(2K - k, \omega^{(m)}_\pm)}$,
such that \begin{align}
           \mathcal{G}\ket{\psi(k, \omega^{(m)}_\pm)} =
\ket{\psi(k, \omega^{(m)}_\mp)} =
 \ket{\psi(2K - k, \omega^{(m)}_\pm)}, \label{parityeq}
          \end{align}
i.e. both states share the same eigenfrequency $\omega^{(m)}_\pm$. 
In terms of the matrix elements of the evolution operator $\hat{U}^{(m)}$ this symmetry implies
\begin{align}
u^{(m)}_{11}(2 K - k) & = \left[ u^{(m)}_{11}(k)\right]^*, \label{u11gp} \\
u^{(m)}_{12}(2K - k) & = e^{i G} \left[ u^{(m)}_{12}(k)\right]^*~. \label{u12gp}
\end{align}
The consequence of (\ref{parityeq}) is $M_s(\omega)$ = $0$. Indeed, the parity operator $\mathcal{G}$ swaps the
spin components and therefore reverts the sign of the magnetization (\ref{Mk}), which then leads to opposite magnetizations of
$\ket{\psi(k, \omega^{(m)}_\pm)} $ and $\ket{\psi(2K - k, \omega^{(m)}_\pm)}$. In operator form the
generalized parity symmetry (\ref{parityeq}),(\ref{u11gp}-\ref{u12gp}) can be expressed as 
\begin{equation}
(\mathcal{P} \otimes \mathcal{G}) \cdot \hat{M} \cdot (\mathcal{P} \otimes \mathcal{G})^\dagger = -\hat{M}\;.
\label{OPgp}
\end{equation}

In order to realize the spectral magnetization ratchet effect, 
i.e.~$M_s(\omega) \neq 0$, one needs to {\it break} the generalized parity symmetry Eq.~(\ref{parityeq}-\ref{OPgp}).
In the next section we explicitly show how to break the generalized parity symmetry and realize the ratchet effect for $m =2, 3$-periodic DTQWs.

\section{The spectral magnetization ratchet}

\subsection{$m=1$}\label{one-step-dispersion}

\begin{figure}[!h]
\includegraphics[width=0.45\textwidth]{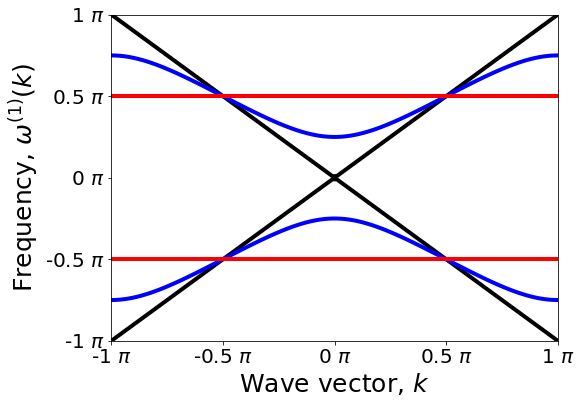}
\caption{
Dispersion relation 
$\omega_{\pm}(k)$ for $m=1$.
$\theta_1 = 0$ (black solid lines),
$\theta_1 = \pi/4$ (blue (light gray) solid lines) 
and $\theta_1 = \pi/2$ (red (gray) solid lines).
Other parameters: $\varphi_{1,1} =\varphi_{2,1}= 0$.
}
\label{fig1}
\end{figure}

For $m=1$, 
the dispersion relation is written explicitly as \cite{vakulchyk2017anderson}
\begin{eqnarray}\label{GeneralDispRel-1step}
  \omega^{(1)}_{\pm}(k)=\pm  \arccos[\cos(\theta_1) \cos (k-\varphi_{1,1})].
\end{eqnarray}
Typical two-band dispersion relations $\omega^{(1)}_\pm(k)$ for various values of $\theta_1$ are shown 
in Fig.~\ref{fig1}.

With (\ref{U-operator}) we find
\begin{eqnarray}
u_{11}^{(1)} = e^{i(\varphi_{1,1}-k)}\cos\theta_1 \;,
\label{m=1u11}
\\
u_{12}^{(1)} = e^{i(\varphi_{2,1}-k)}\sin\theta_1 \;.
\label{m=1u12}
\end{eqnarray}
It follows that the generalized parity symmetry relations (\ref{u11gp},\ref{u12gp}) are satisfied for all coin parameters of the DTQW
with the notations 
$K = \varphi_{1,1}$ and $G = 2(\varphi_{2,1} - \varphi_{1,1})$. 
Therefore the spectral magnetization $M_s(\omega)=0$, and a single period DTQW always possesses generalized
parity symmetry. This happens remarkably despite the action of both nonzero external and internal magnetic flux $\varphi_{1,1}$ and 
$\varphi_{2,1}$.

\begin{figure}[!h]
\includegraphics[width=0.4\textwidth]{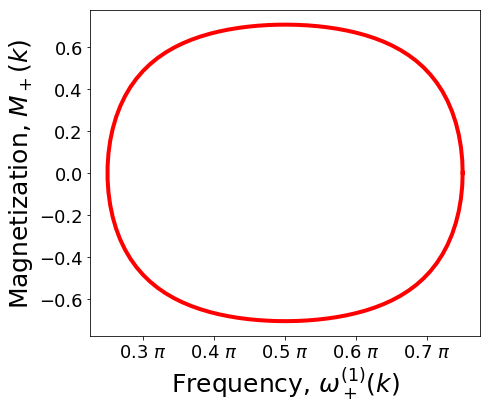}
\caption{Typical dependence of $M_+(k)$ from Eq.(\ref{magnetization-1step}) which forms a symmetric loop around $M=0$. The $x$-axis values of the frequency are obtained from
Eq.(\ref{GeneralDispRel-1step}).
The spectral magnetization $M_s(\omega)=0$. Here $\theta_1 = \pi/4$ and $\varphi_{1,1} = \varphi_{2,1}=0$.
}
\label{fig5}
\end{figure}

For the one-periodic DTQW all eigenstates are doubly degenerated, and using Eqs.~(\ref{GeneralDispRel-1step}) and (\ref{Magnetization-Gen-1}) we obtain 
\begin{eqnarray}\label{magnetization-1step}
 M_{+}(k)  =  \frac{\cos(\theta_1) \sin(k - \varphi_{1,1})}{\sqrt{1 - \cos^2(\theta_1) \cos^2(k - \varphi_{1,1})}} \;.
\end{eqnarray}
In line with the above symmetry analysis the spectral magnetization vanishes, as also observed from the loop symmetry in Fig.\ref{fig5}.

\subsection{$m=2$}
\begin{figure}[!h]
\includegraphics[width=0.45\textwidth]{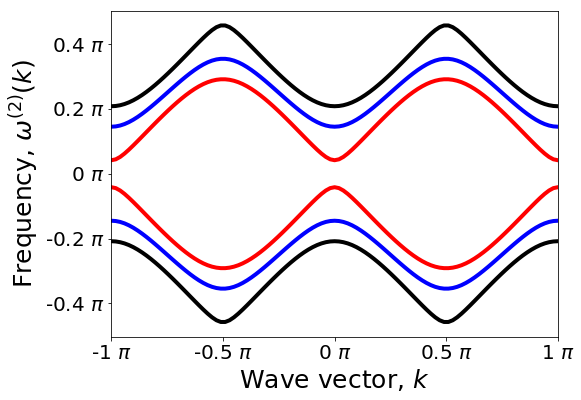}
\caption{Dispersion relation $\omega_{\pm}(k)$ for $m=2$. 
$\varphi_{2,1}=0$ (black solid lines); $\varphi_{2,1}=\pi/2$ (blue (light gray) line); 
$\varphi_{2,1}=\pi$ (red (gray) line). The other parameters are fixed to $\theta_{1}=\pi/4$, $\theta_{2}=\pi/6$,
$\varphi_{1,1}=\varphi_{1,2}= \varphi_{2,2} = 0$.
}\label{fig3}
\end{figure}

For $m=2$ we
calculate  the product of two operators $\hat U_1(k)$ and $\hat U_2(k)$ and find
\begin{multline}\label{2_elements}
 u^{(2)}_{11} = \cos(\theta_1) \cos(\theta_2) e^{-i(2k-\varphi_{1,1}-\varphi_{1,2})}\\
  -\sin(\theta_1) \sin(\theta_2) e^{i(\varphi_{2,2}-\varphi_{2,1})},~~~~~~~~~~~~~~\\
 u^{(2)}_{12} = \sin(\theta_1) \cos(\theta_2) e^{-i(2k - \varphi_{1,2} - \varphi_{2,1})}~~~~~~~~~~~~~~~~~~~\\
 + \cos(\theta_1) \sin(\theta_2) e^{-i(\varphi_{1,1} - \varphi_{2,2})}\;.~~~~~~~~~~~~~~
\end{multline}
With the help of Eqs.~(\ref{Operator})-(\ref{GeneralDispRel}) we obtain the explicit expression for the dispersion relation $\omega^{(2)}_{\pm}(k)$ as
\begin{multline}\label{GeneralDispRel-2step}
 \omega^{(2)}_{\pm}(k)=\pm \frac{1}{2}\arccos[\cos(\theta_1) \cos(\theta_2) \cos (2k-\varphi_{1,1}-\varphi_{1,2})\\
  -\sin(\theta_1) \sin(\theta_2) \cos (\varphi_{2,1}-\varphi_{2,2})]~.~~~~
\end{multline}
Typical band structures are shown in Fig.\ref{fig3}.

In order to possess generalized parity symmetry (\ref{u11gp},\ref{u12gp}), it follows from Eq.(\ref{2_elements}) that
$\sin(\theta_1) \sin(\theta_2) e^{i(\varphi_{2,2}-\varphi_{2,1})}=0$. 
Then it follows that 
\begin{equation}
K = \frac{\varphi_{1,1}+\varphi_{1,2}}{2} \;.
\label{m2K}
\end{equation}
Three symmetry cases can be distinguished.
\begin{eqnarray}
\label{m2i}
S_{2,1} \;: \;  \theta_1=n \pi \rightarrow  G = 2(\varphi_{2,2} - \varphi_{1,1})\;,
\\
S_{2,2} \; : \; \theta_2=n \pi \rightarrow  G = -2( \varphi_{2,1}  + \varphi_{1,1})\;,
\label{m2ii}
\\
S_{2,3} \; : \; \varphi_{2,2} - \varphi_{2,1} = n\pi \rightarrow  G = -2( \varphi_{2,1}  + \varphi_{1,1}).
\label{m2iii}
\end{eqnarray}
Here $n=0,\pm1,\pm2,...$ is an arbitrary integer. If all of the above conditions are broken, then we can expect a nonzero
spectral magnetization ratchet to appear. If on the contrary at least one of the above symmetry conditions $S_{2,1},S_{2,2},S_{2,3}$ is satisfied, the
spectral magnetization vanishes for all frequencies.

For the two-periodic DTQW the eigenstates are $4$-fold degenerated, and using Eqs.~(\ref{2_elements}) and (\ref{GeneralDispRel-2step}) we obtain
%\begin{widetext}
\begin{align}\label{magnetization-2step}
& M_+(k) \nonumber\\
= & \frac{\cos(\theta_1) \cos(\theta_2) \sin(2k - 2 k_0) 
- \sin(\theta_1) \sin(\theta_2) \sin(\delta\varphi_{2})}{|\sin(2 \omega^{(2)}_+(k))|}.
\end{align}
%\end{widetext}
where $\delta \varphi_{2}=\varphi_{2,1}-\varphi_{2,2}$, $k_0 = (\varphi_{1,1} + \varphi_{1,2})/2$.
\begin{figure}[!h]
\subfigure[]{\includegraphics[width=0.45\textwidth]{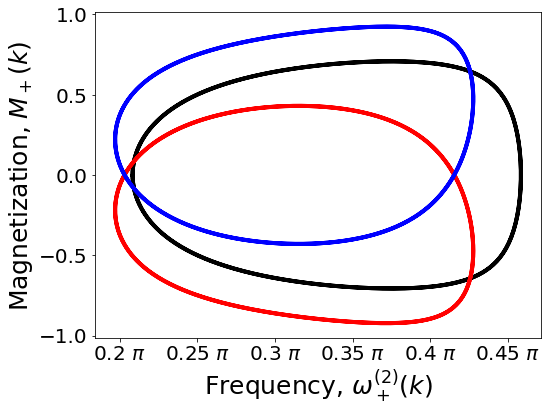}}
\subfigure[]{\includegraphics[width=0.45\textwidth]{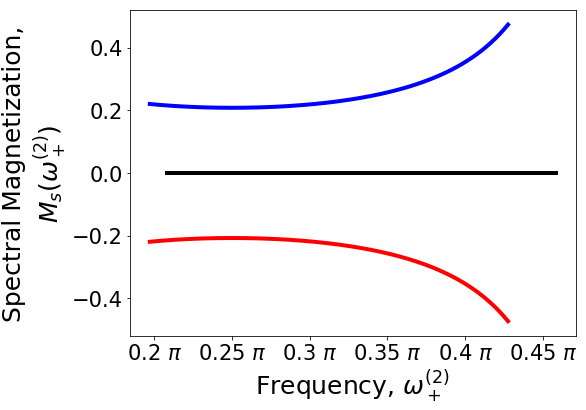}}
\caption{(a) Typical dependencies of $M_+(k)$ versus $\omega_+^{(2)}(k)$.
Black line (symmetric case) 
$\delta \varphi_2=0$; blue (light gray) line (non-symmetric case) $\delta \varphi_2=\pi/5$; red (gray) line (non-symmetric case), 
$\delta \varphi_2=-\pi/5$. Here  $\theta_1=\pi/4$, $\theta_2=\pi/6$ and $\varphi_{1,1}=\varphi_{1,2}=0$.
(b) Spectral magnetization $M_s(\omega)$ for the corresponding plots of $M_+(k)$ from (a).
}
\label{fig6}
\end{figure}
The typical dependencies of $M_+(k)$ for different values of phase shift  
$\delta \varphi_{2}$ are shown in Fig.~\ref{fig6}.
Nonzero spectral magnetization values appear once $\delta \varphi_{2} \neq 0$, signaling the breaking
of generalized parity symmetry (\ref{m2i}-\ref{m2iii}) and the appearance of the spectral magnetization ratchet.

\subsection{$m=3$}

For the $3$-periodic DTQW we
calculate the product of three operators $\hat U_1(k)$, $\hat U_2(k)$, $\hat U_3(k)$
and get 
\begin{align}\label{first_element_3}
 u^{(3)}_{11}(k) = &\cos (\theta_1)\cos(\theta_2) \cos(\theta_3)   e^{i (k_a - 3 k)}\nonumber\\
 &- \sin(\theta_1)\sin(\theta_2)\cos(\theta_3)  e^{i (k_b -  k)}\nonumber\\
& - \sin (\theta_1) \cos (\theta_2) \sin (\theta_3) e^{i ( -k_c  +  k)} \nonumber\\
& -  \cos (\theta_1) \sin (\theta_2) \sin (\theta_3) e^{i ( k_d   -  k)}
\end{align}
with the notations
\begin{eqnarray}
k_{a} = \varphi_{1,1} + \varphi_{1,2} + \varphi_{1,3},
\\
k_b = \varphi_{1,3} - \varphi_{2,1} + \varphi_{2,2},
\\
k_c = \varphi_{1,2}       +\varphi_{2,1}    -\varphi_{2,3},
\\
k_d = \varphi_{1,1}-\varphi_{2,2}+\varphi_{2,3} .
\end{eqnarray}
Note that $k_a = k_b + k_c + k_d$.
The off-diagonal element follows as
\begin{align}\label{second_element_3}
u^{(3)}_{12}(k)  =  &\sin(\theta_1) \cos (\theta_2)\cos(\theta_3) e^{i(k_e  - 3 k)}\nonumber\\
&+ \cos(\theta_1) \sin(\theta_2) \cos(\theta_3)  e^{i (k_f -  k)}\nonumber\\
&+ \cos(\theta_1) \cos(\theta_2) \sin(\theta_3) e^{i (-k_g  +  k)}\nonumber \\
&-  \sin(\theta_1)\sin(\theta_2)\sin(\theta_3)  e^{i (k_h - k)} ~.
\end{align}
with the notations
\begin{eqnarray}
k_{e} =  &  \varphi_{1,2} +  \varphi_{1,3} +  \varphi_{2,1} = k_a + \varphi_{2,1} - \varphi_{1,1},
\\
k_f  = &  - \varphi_{1,1} + \varphi_{1,3}  +    \varphi_{2,2} = k_b + \varphi_{2,1} - \varphi_{1,1},
\\
k_g = & \varphi_{1,1} + \varphi_{1,2} - \varphi_{2,3} = k_c - \varphi_{2,1} + \varphi_{1,1},
\\
k_h = & \varphi_{2,1} - \varphi_{2,2} +  \varphi_{2,3} = k_d + \varphi_{2,1} - \varphi_{1,1}.
\end{eqnarray}

\begin{figure}[!h]
\includegraphics[width=0.45\textwidth]{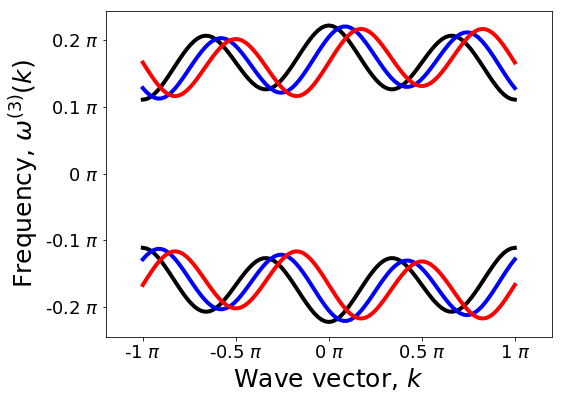}
\caption{
Dispersion relation $\omega_{\pm}(k)$ for the three-periodic ($m=3$)
DTQW with different angles $\varphi_{1,1}$: 
 symmetric case $\varphi_{1,1}=0$ (black solid line);  non-symmetric cases $\varphi_{1,1}=\pi/4$ (blue (light gray) line), $\varphi_{1,1}=\pi/2$  (red (gray) line). 
Here $\theta_{1}=\pi/3$, $\theta_{2}=\pi-0.43$, $\theta_{3}=0.43$ are chosen. All other angles set to zero.}
\label{fig4}
\end{figure}

Using Eqs.~(\ref{Operator})-(\ref{GeneralDispRel}) 
we obtain the explicit expression for the dispersion relation $\omega^{(3)}_{\pm}(k)$ as
\begin{multline}\label{GeneralDispRel-3step}
 \omega^{(3)}_{\pm}(k)=\pm \frac{1}{3} \arccos[\cos(\theta_1) \cos(\theta_2)\cos(\theta_3) \cos (3k-k_a)\\
 -\sin(\theta_1) \sin(\theta_2)\cos(\theta_3) \cos (k-k_b)\\
 -\sin(\theta_1) \cos(\theta_2)\sin(\theta_3) \cos (k-k_c)\\
 -\cos(\theta_1) \sin(\theta_2)\sin(\theta_3) \cos (k-k_d)]~.~~~
\end{multline}
Typical band structures are shown in Fig.\ref{fig4}.

Let us identify parameters for which the generalized parity symmetry holds. We distinguish two symmetry conditions - $S_{3,1}$ and $S_{3,2}$.
$S_{3,1}$ constrains the coin parameters $\theta_i$, while leaving all other angles 
arbitrary:
\begin{equation}
S_{3,1} \; : \; \theta_i=n\pi/2 \;, \theta_{j \neq i}=m\pi/2
\label{s31}
\end{equation}
for arbitrary integers $n,m$. The details of the cumbersome analysis, including the values of $K$ and $G$ are outsourced to Appendix \ref{appen_1}.

The second generalized parity symmetry case $S_{3,2}$ constrains all but the coin parameters $\theta_i$. It is realized when 
$k_a=3k_b=3k_c=3k_d$ which implies $k_b=k_c=k_d$. These conditions reduce to 
\begin{equation}
S_{3,2} \; : \; \Bigg\{
\begin{array}{c}
\varphi_{2,2}  =  \frac{1}{3}( \varphi_{1,1} + \varphi_{1,2} - 2\varphi_{1,3})+\varphi_{2,1}
\\ \\
 \varphi_{2,3}   =   \frac{1}{3}( -\varphi_{1,1} + 2\varphi_{1,2} - \varphi_{1,3})+\varphi_{2,1}\
\end{array}
\end{equation}
with the parameters of the generalized parity symmetry reading
\begin{equation}
K=k_b \;,\; G = 2(\varphi_{2,1} - \varphi_{1,1}) \;.
\end{equation}
If any of the two symmetries $S_{3,1}$ and $S_{3,2}$ holds, the spectral magnetization vanishes. If both are violated, a nonzero spectral magnetization ratchet is predicted.

Using Eqs.~(\ref{first_element_3}), (\ref{second_element_3}), (\ref{GeneralDispRel-3step}) and (\ref{Magnetization-Gen-1})
we obtain the explicit expression for $M_+(k)$:
\begin{figure}[!h]
\subfigure[]{\includegraphics[width=0.45\textwidth]{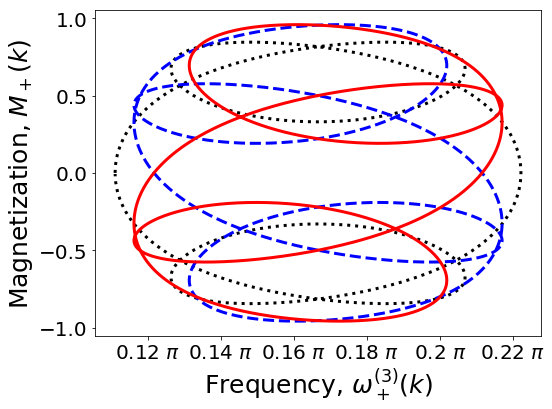}}
\subfigure[]{\includegraphics[width=0.45\textwidth]{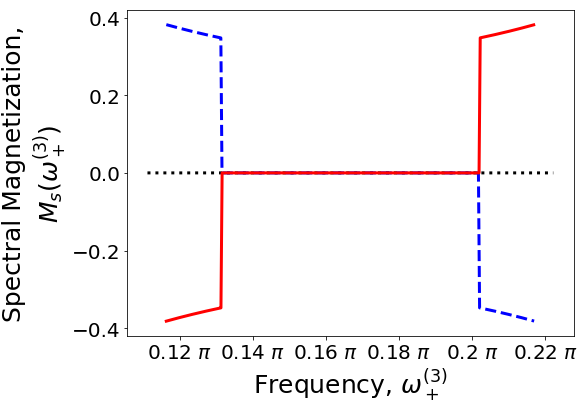}}
\caption{(a) $M^{(3)}_+(k)$ as a function of frequency $\omega_+^{(3)}(k)$
for different values of $\delta \varphi_{1,1}$:  black dotted line (symmetric case) $\varphi_{1,1}=0$; 
blue (light gray) dashed line (non-symmetric case) $\varphi_{1,1}=\pi/2$;  red (gray)
solid line (non-symmetric case) $\varphi_{1,1}=-\pi/2$; 
Here $\theta_{1}=\pi/3$, $\theta_{2}=\pi-0.43$, $\theta_{3}=0.43$, and $\varphi_{1,2}=\varphi_{1,3}=\varphi_{2,1}=\varphi_{2,2}=\varphi_{2,3}=0$.
(b) Spectral magnetization $M_s(\omega)$ for the corresponding plots of $M_+(k)$ in (a).
}
\label{fig7}
\end{figure}

 \begin{align}
 M_+(k)
 = \Big[\cos(\theta_1) \cos(\theta_2)\cos(\theta_3) \sin(3k-k_a) \nonumber\\
  -\sin(\theta_1) \sin(\theta_2)\cos(\theta_3) \sin (k-k_b) \nonumber\\
  + \sin(\theta_1) \cos(\theta_2)\sin(\theta_3) \sin(k-k_c) \nonumber\\
 -\cos(\theta_1) \sin(\theta_2)\sin(\theta_3) \sin(k-k_d) \Big]\frac{1}{|\sin(3\omega_+^{(3)}(k))|} ~.
\end{align}
We consider a case where all $\varphi_{i,j} = 0$ except $\varphi_{1,1}$, and $\theta_i \neq n\pi/2$ for any $i$ and any integer $n$.
The corresponding dispersion relation for such a case is shown in Fig. \ref{fig4}. From the previous analysis, 
it follows that the spectral magnetization must vanish if $\varphi_{1,1}=0$ since then $S_{3,2}$ is restored.
The dependence $M_+(k)$ for different values of the angle $\varphi_{1,1}$ is shown in Fig.~\ref{fig7}a. 
Indeed the spectral magnetization $M_s(\omega)=0$ is obtained if $\varphi_{1,1}=0$  (see,  black dotted line in Fig.~\ref{fig7}b),
as a direct consequence of the generalized parity symmetry with parameters $K = 0$, $G = 0$. For nonzero values
of $\varphi_{1,1}$, the parity symmetry as described by Eq.~(\ref{PundG}) is broken, and non-zero values of spectral magnetization are obtained 
(see blue (light gray) and red (gray) line in Fig.~\ref{fig7}).

\section{Quantum measurements of the spectral magnetization ratchet effect}

Let us discuss ways to observe the spectral magnetization $M_s(\omega)$
in the quantum evolution of $m$-periodic DTQWs.
We introduce the time-dependent correlation function 
\begin{align} \label{Corrfunction}
C_M(t,\tau) = \sum_n \psi_{+}(n,t)\psi^\star_{+}(n,t-\tau) - \psi_{-}(n,t)\psi^\star_{-}(n,t-\tau),
\end{align} where we consider the time-steps $t$, $\tau$ as multiples of $m$. 
Its discrete Fourier-transformation w.r.t.~$\tau$ with additional averaging over the discrete time $t$ 
\begin{align}
 C_M(\omega) = \sum_{\tau = 0}^\infty e^{i\omega \tau}
 \left[\lim_{T \to \infty} \frac{1}{T} \sum_{t = 0}^{T} C_M(t,\tau)\right]
\end{align}
can be expressed as 
\begin{equation} \label{Corrfunction-Fourier}
C_M(\omega)=\sum_{l} |\alpha (k_l, \omega)|^2 M(k_l),
\end{equation}
where the index $l$ in this sum runs over all degenerate points corresponding to the frequency 
$\omega$. 
Note here that $T$ is also a multiple of $m$.
The coefficients $\alpha(k_l, \omega)$ are determined by the initial conditions:
 \begin{align}
 \psi_\pm(n, t = 0) =  \frac{1}{\sqrt{N}} \sum_{k} \sum_{p = \pm}  e^{i k n} \alpha (k, \omega_p) \psi_\pm(k, \omega_p)~.
 \end{align}
Assuming a homogeneous distribution of such coefficients $\alpha (k, \omega)$ such that $|\alpha (k, \omega)|^2 = const$, 
we obtain $C_M(\omega) \propto M_s(\omega)$  (see derivation details in Appendix \ref{appen_2}).
Note that the assumption of all basis states having the same weight is similar to a generalized notion of infinite temperature. 
It follows that the infinite temperature states of quantum Floquet systems like in the case of $m$-periodic DTQWs may keep a nontrivial internal structure
characterized by the presence or absence of certain symmetries.

The correlator $C_M(\omega)$ can be directly measured using a continuous quantum measurements setup proposed and regularly used for the study of quantum dynamics of superconducting qubit networks
\cite{wallraff2004strong,volkov2014collective,macha2014implementation}. 
These setups consist of a low-dissipative transmission line weakly coupled with the studied quantum system (here the DTQW). The transmission line is 
characterized by a discrete set of internal mode frequencies at which the transmission coefficient is suppressed. Let us consider one such mode with frequency 
$\omega_0$. Due to the additional coupling of the line with the DTQW 
%
%In such a setup the transmission coefficient $D(\omega)$ of electromagnetic waves propagating in a low-dissipative waveguide weakly coupled to the studied quantum system (e.g. the DTQW),  
%is measured. T
the transmission coefficient $D(\omega)$ will display a resonant drop at the resonant frequency $\omega_{res}$:
\begin{equation}
D(\omega)=1- \frac{\alpha}{(\omega-\omega_{res})^2+\gamma^2},
\label{resonantdrops}
\end{equation}
where $\alpha$ is the strength of the resonance and $\gamma \ll \omega$ is the dissipation parameter. The location of the resonance $\omega_{res}$ is renormalized due to the presence of the weakly coupled DTQW: 
%depending on the value of $C_M(\omega_{res})$ as
$\omega_{res}=\omega_0+\chi C_M(\omega_{0})$, 
%where $\omega_0$ is the resonant frequency of the waveguide decoupled from the studied quantum system, 
where $\chi$ is determined by the small coupling strength between the waveguide and the DTQW. Therefore, this method  
allows to measure the value of $C_M(\omega)$, and consequently allows to observe the predicted appearance of the spectral magnetization ratchet.

\section{Conclusion}\label{conclud}

We have shown that a spectral magnetization ratchet can be observed in a spatially homogeneous DTQW system by breaking a generalized parity symmetry. 
To achieve that goal, we need to introduce a generalized discrete time quantum walk process with quantum coins varying
periodically in time. As a result, we obtained conditions for the generalized parity symmetry to hold for $m=2$ and $m=3$,
and identified systematic ways to break this symmetry by proper parameter choices. As a result, 
a non-vanishing spectral magnetization is obtained, which tells that a resonant excitation of all (degenerate) eigenstates at a given eigenfrequency
$\omega$ will lead to a non-vanishing population imbalance, or simply magnetization. 
Our results  add new possibilities to the control of quantum networks in quantum simulation setups using
methods developed in condensed matter physics.

\section*{Acknowledgment}
This work was supported by the Institute for Basic Science, Project Code (IBS-R024-D1).  
P. K. thanks the hospitality of the Center for Theoretical Physics of Complex Systems and the Korean Undergraduate Science Program KUSP2019 (kusp.ibs.re.kr) at the Institute for Basic Science for hospitality and financial support.
M. V. F. 
thanks the partial financial support of Ministry of Science and Higher Education of the Russian
Federation in the framework of Increase Competitiveness Program of NUST 'MISiS' K2-2017-081. 

%\bibliography{Ihor_QW,Ratchet}

\appendix 

\section{Symmetry $S_{3,1}$ for $m=3$}\label{appen_1}
Here we analyse the generalized parity symmetry $S_{3,1}$ which holds when $\theta_i=n\pi/2$ and  $\theta_j=m\pi/2$ is satisfied for any pair of $i\neq j$.
In the following
$n$, $m$ are arbitrary integers.
\subsection{$i=1,j=2$}
\begin{itemize}

 \item For $\theta_1 = (2n+1)\frac{\pi}{2}$, $\theta_2 = m \pi$, we have 
 \begin{align}
 & u^{(3)}_{11}(k)  =  - (-1)^{m+n}\sin (\theta_3) e^{i ( -k_c  +  k)},\nonumber\\
  & u^{(3)}_{12}(k) =  (-1)^{m+n}\cos(\theta_3) e^{i(k_e  - 3 k)}\nonumber\\
 &\Rightarrow u^{(3)}_{11}(2 k_c - k) = \left[u^{(3)}_{11}(k)\right]^*,~ \nonumber\\
 &u^{(3)}_{12}(2 k_c - k) = e^{2 i(k_e - 3 k_c)} \left[u^{(3)}_{12}(k)\right]^*.
\end{align}
 \begin{align}
   \Rightarrow K = k_c, G = 2(k_e - 3k_c). 
   \end{align}

\item For $\theta_1 = n \pi$, $\theta_2 = m \pi$, we have 

\begin{align}
 u^{(3)}_{11}(k) = (-1)^{m+n}\cos(\theta_3)   e^{i (k_a - 3 k)},\nonumber\\
 u^{(3)}_{12}(k)  =  (-1)^{m+n} \sin(\theta_3) e^{i (-k_g +  k)} \nonumber\\
 \Rightarrow  u^{(3)}_{11}(2k_a/3 - k) = \left[u^{(3)}_{11}(k)\right]^*,\nonumber\\
 u^{(3)}_{12}(2k_a/3 - k) = e^{2i k_a/3 - 2i k_g} \left[u^{(3)}_{12}(k)\right]^*,
\end{align}
\begin{align}
 \Rightarrow K = k_a/3,~ G = 2 k_a/3 - 2 k_g~.
\end{align}

\item For $\theta_1 = n \pi$, $\theta_2 = (2m+1)\frac{\pi}{2}$, we have 
\begin{align}
 u^{(3)}_{11}(k) =  - (-1)^{m+n} \sin (\theta_3) e^{i ( k_d   -  k)},\nonumber\\
 u^{(3)}_{12}(k)  =   (-1)^{m+n} \cos(\theta_3)  e^{i (k_f -  k)} \nonumber\\
 \Rightarrow u^{(3)}_{11}(2k_d - k) = \left[ u^{(3)}_{11}(k)\right]^*,\nonumber\\
 u^{(3)}_{12}(2k_d - k) = e^{2i(k_f - k_d)}\left[ u^{(3)}_{12}(k)\right]^*~.
\end{align}
\begin{align}
 \Rightarrow K = k_d, G = 2(k_f - k_d)~.
\end{align}

\item  For $\theta_1 = (2n+1)\frac{\pi}{2}$, $\theta_2 = (2m +1) \frac{\pi}{2}$, we have 
\begin{align}
 u^{(3)}_{11}(k) =  - (-1)^{m+n}\cos(\theta_3)  e^{i (k_b -  k)},\nonumber\\
 u^{(3)}_{12}(k)  =  - (-1)^{m+n}\sin(\theta_3)  e^{i (k_h - k)}, \nonumber\\
 \Rightarrow  u^{(3)}_{11}(2k_b - k) = \left[ u^{(3)}_{11}(k)\right]^*,\nonumber\\
 u^{(3)}_{12}(2k_b - k) = e^{2i(k_h - k_b)}\left[ u^{(3)}_{12}(k)\right]^*~.
\end{align}
\begin{align}
\Rightarrow  K  = k_b, G = 2(k_h - k_b)~.
\end{align}

\end{itemize}

\subsection{$i=1,j=3$}

\begin{itemize}
\item For $\theta_1 = (2n+1)\frac{\pi}{2}$, $\theta_3 = m \pi$, we have 
\begin{align}
 u^{(3)}_{11}(k) =  - (-1)^{m+n}\sin(\theta_2)  e^{i (k_b -  k)},\nonumber\\
 u^{(3)}_{12}(k)  =  (-1)^{m+n} \cos (\theta_2) e^{i(k_e  - 3 k)},\nonumber\\
 \Rightarrow  u^{(3)}_{11}(2k_b - k) = \left[  u^{(3)}_{11}(k)\right]^*, \nonumber\\
  u^{(3)}_{12}(2k_b - k) = e^{2i(k_e - 3k_b)}\left[  u^{(3)}_{12}(k)\right]^*
 \end{align} 
 \begin{align}
  \Rightarrow K = k_b, G = 2(k_e - 3k_b)~.
 \end{align}

\item For $\theta_1 = n \pi$, $\theta_3 = m \pi$, we have 

\begin{align}
 u^{(3)}_{11}(k) = (-1)^{m+n}\cos(\theta_2) e^{i (k_a - 3 k)},\nonumber\\
 u^{(3)}_{12}(k)  =   (-1)^{m+n} \sin(\theta_2) e^{i (k_f -  k)}~.\nonumber\\
 \Rightarrow u^{(3)}_{11}(2k_a/3 - k) = \left[u^{(3)}_{11}(k) \right]^*,\nonumber\\
 u^{(3)}_{12}(2k_a/3 - k) = e^{2ik_f - 2 i k_a/3}\left[u^{(3)}_{12}(k) \right]^*~.
\end{align}
\begin{align}
 K  = k_a/3, G = 2k_f - 2k_a/3~.
\end{align}

\item For $\theta_1 = n \pi$, $\theta_3 = (2m+1)\frac{\pi}{2}$, we have 
\begin{align}
 u^{(3)}_{11}(k) = - (-1)^{m+n} \sin (\theta_2) e^{i (k_d   -  k)},\nonumber\\
 u^{(3)}_{12}(k)  =  (-1)^{m+n} \cos(\theta_2) e^{i (-k_g +  k)},\nonumber\\
 \Rightarrow u^{(3)}_{11}(2k_d - k) = \left[ u^{(3)}_{11}(k)\right]^*,\nonumber\\
 u^{(3)}_{12}(2k_d - k) = e^{-2i(k_g - k_d)}\left[ u^{(3)}_{12}(k)\right]^*.
\end{align}
\begin{align}
 K  = k_d, G = -2(k_g - k_d)~.
\end{align}

\item  For $\theta_1 = (2n+1)\frac{\pi}{2}$, $\theta_3 = (2m +1) \frac{\pi}{2}$, we have 
\begin{align}
 u^{(3)}_{11}(k) =  - (-1)^{m+n} \cos (\theta_2) e^{i ( -k_c  +  k)},\nonumber\\
 u^{(3)}_{12}(k)  = - (-1)^{m+n} \sin(\theta_2)  e^{i (k_h - k)} \nonumber\\
 \Rightarrow  u^{(3)}_{11}(2k_c - k) = \left[ u^{(3)}_{11}(k)\right]^*,\nonumber\\
  u^{(3)}_{12}(2k_c - k) = e^{2i(k_h - k_c)}\left[ u^{(3)}_{12}(k)\right]^*
 \end{align}
\begin{align}
 K = k_c, G = 2(k_h - k_c)~.
\end{align}

\end{itemize}

\subsection{$i=2,j=3$}
\begin{itemize}

\item For $\theta_2 = (2n+1)\frac{\pi}{2}$, $\theta_3 = m \pi$, we have 

\begin{align}
 u^{(3)}_{11}(k) =  - (-1)^{m+n} \sin(\theta_1) e^{i (k_b -  k)},\nonumber\\
u^{(3)}_{12}(k) = (-1)^{m+n} \cos(\theta_1) e^{i (k_f -  k)},\nonumber\\
\Rightarrow u^{(3)}_{11}(2k_b - k) = \left[ u^{(3)}_{11}(k) \right]^*,\nonumber\\
u^{(3)}_{12}(2k_b - k) = e^{2i(k_f - k_b)}\left[ u^{(3)}_{12}(k) \right]^*~.
 \end{align}
\begin{align}
 K = k_b, G = 2(k_f - k_b)~.
\end{align}

\item For $\theta_2 = n \pi$, $\theta_3 = m \pi$, we have 
\begin{align}
 u^{(3)}_{11}(k) = (-1)^{m+n}\cos (\theta_1) e^{i (k_a - 3 k)},\nonumber\\
 u^{(3)}_{12}(k)  = (-1)^{m+n} \sin(\theta_1) e^{i(k_e  - 3 k)}, \nonumber\\
 \Rightarrow u^{(3)}_{11}(2k_a/3 - k) = \left[ u^{(3)}_{11}(k)\right]^*,\nonumber\\
 u^{(3)}_{12}(2k_a/3 - k) = e^{2i(k_e - k_a)}\left[ u^{(3)}_{12}(k)\right]^*~.
\end{align}
 \begin{align}
  K  = k_a/3, G  = 2(k_e - k_a)~.
 \end{align}

 \item For $\theta_2 = n \pi$, $\theta_3 = (2m+1)\frac{\pi}{2}$, we have 
\begin{align}
 u^{(3)}_{11}(k) = - (-1)^{m+n} \sin (\theta_1) e^{i ( -k_c  +  k)},\nonumber\\
 u^{(3)}_{12}(k)  = (-1)^{m+n}  \cos(\theta_1) e^{i (-k_g +  k)},\nonumber\\
 \Rightarrow  u^{(3)}_{11}(2k_c - k) = \left[  u^{(3)}_{11}(k)\right]^*,\nonumber\\
 u^{(3)}_{12}(2k_c - k) = e^{2i(k_c - k_g)}\left[u^{(3)}_{12}(k)\right]^*.
\end{align}
\begin{align}
 K  = k_c,~G  = 2(k_c - k_g)~.
\end{align}

\item  For $\theta_2= (2n+1)\frac{\pi}{2}$, $\theta_3 = (2m +1) \frac{\pi}{2}$, we have 
\begin{align}
 u^{(3)}_{11}(k) =  - (-1)^{m+n} \cos (\theta_1) e^{i ( k_d   -  k)},\nonumber\\
 u^{(3)}_{12}(k)  =  - (-1)^{m+n} \sin(\theta_1) e^{i (k_h - k)}, \nonumber\\
 \Rightarrow u^{(3)}_{11}(2k_d - k) = \left[ u^{(3)}_{11}(k) \right]^*,\nonumber\\
 u^{(3)}_{11}(2k_d - k) = e^{2i(k_h - k_d)}\left[ u^{(3)}_{11}(k) \right]^*~.
\end{align}
\begin{align}
 K  = k_d, G  = 2(k_h - k_d)~.
\end{align}

\end{itemize}

\section{Derivation of the relation between the discrete time-dependent correlation function and the spectral magnetization}\label{appen_2}
We consider all  time intervals as multiples of  $m$, so that $t$, $\tau$ $\in$ $m \mathbb{Z}$.  

The expectation value of the operator: $\left(U^{(m)}\right)^{\tau/m} \cdot \hat{M}$ w.r.t.~a general state $\ket{\psi(t)}$ at time-step $t$ can be written as
\begin{align}\label{eq1}
 \text{Tr}\big[\ket{\psi(t)}\bra{\psi(t)} \cdot \left(U^{(m)}\right)^{\tau/m} \cdot \hat{M}\big] \hspace{1.8cm} &\nonumber\\
 =  \text{Tr}\left[ \ket{\psi(t)}\bra{\psi(t - \tau)} \cdot \hat{M}\right] \hspace{3.3cm} &  \nonumber\\
 = \sum_n \left\{ \bra{n} \otimes \bra{+}~ \ket{\psi(t)}\bra{\psi(t - \tau)} \cdot \hat{M}\ket{n} \otimes \ket{+} \right\}  \nonumber\\
 +  \left\{ \bra{n}  \otimes \bra{-}~ \ket{\psi(t)}\bra{\psi(t - \tau)} \cdot \hat{M}\ket{n} \otimes \ket{-} \right\} \nonumber\\
 =   \sum_n \psi_+(n,t) \psi^*_+(n,t-\tau) - \psi_-(n,t) \psi^*_-(n,t-\tau). 
\end{align}
The last expression in the Eq.~(\ref{eq1}) is denoted by $C_M(t,\tau)$ in the main text. 

We write a general initial state as a superposition of orthogonal basis states (composite states of momentum basis and coin basis):
\begin{align}
 \ket{\psi(t = 0)}
 = \sum_k \sum_{p = \pm}  \alpha(k,\omega_p(k)) \ket{k} \otimes \ket{\psi(k, \omega_p(k))} 
 \end{align}
 \begin{align}
 \Rightarrow & ~\psi_\pm(n,t=0) 
 = \braket{n, \pm| \psi(t = 0)} \nonumber\\ 
 =&   \sum_k \sum_{p = \pm}  \braket{n|k} \alpha(k,\omega_p(k))  \psi_\pm(k, \omega_p(k))\nonumber\\
 =& \frac{1}{\sqrt{N}} \sum_k \sum_{p = \pm} e^{ikn} \Big[\alpha(k,\omega_p(k)) \psi_\pm(k, \omega_p(k)) \Big]
\end{align}
with $N$ being the number of lattice sites, so that 
\begin{align}
 \ket{k} = \frac{1}{\sqrt{N}} \sum_n e^{ikn} \ket{n} \Rightarrow \braket{n|k} = \frac{1}{\sqrt{N}} e^{ikn}.  
\end{align}
The general state at any time-step $t$ is 
\begin{align}
 \psi_\pm(n,t) = \frac{1}{\sqrt{N}} \sum_{k,p} e^{i k n -i \omega_p(k) t} \alpha(k,\omega_p(k)) \psi_\pm(k, \omega_p(k))~.
\end{align}
Therefore \begin{align}
           &\psi_+(n,t) \psi^*_+(n,t - \tau) \nonumber\\
           &= \frac{1}{N} \sum_{k,p} \sum_{k',p'} e^{i(k-k')n} e^{-i \omega_p(k) t} \alpha(k,\omega_p(k)) \psi_+(k, \omega_p(k)) \times \nonumber\\
           & ~~~~~~ e^{i \omega_{p'}(k') (t-\tau)} \alpha^*(k',\omega_{p'}(k')) \psi^*_+(k', \omega_{p'}(k')) ~.\end{align}
 Using  
$\sum_{n = 1}^N e^{i(k-k')n} = N \delta_{kk'}
$
we get 
 \begin{align}\label{eq2}
 \sum_n \psi_+(n,t) \psi^*_+(n,t - \tau) \nonumber\\
  = \sum_k \Big[ \sum_p e^{-i \omega_p(k) t} \alpha(k,\omega_p(k)) \psi_+(k, \omega_p(k))\Big] \times \nonumber\\
  \Big[ \sum_{p'} e^{i \omega_{p'}(k) (t-\tau)} \alpha^*(k,\omega_{p'}(k)) \psi^*_{p'}(k, \omega_+(k))\Big] \nonumber\\ 
  = \sum_k \sum_p e^{- i \omega_p(k) \tau} |\alpha(k,\omega_p(k))|^2 |\psi_+(k, \omega_p(k))|^2 \nonumber\\
  + \sum_k \sum_{p \neq p'} e^{-i \omega_p(k) t} \alpha(k,\omega_p(k)) \psi_+(k, \omega_p(k)) \times \nonumber\\
 e^{i \omega_{p'}(k) (t-\tau)} \alpha^*(k,\omega_{p'}(k)) \psi^*_{+}(k, \omega_{p'}(k))~. 
          \end{align}
Averaging over the time steps $t$ leads to a vanishing of the cross terms
in Eq.~(\ref{eq2})  since
\begin{align}
\lim_{T \rightarrow \infty} 
\frac{1}{T} \sum_{t = 0}^{T} e^{\pm i \left[\omega_-(k) - \omega_+(k)\right] t} = \delta_{\omega_+(k)~\omega_-(k)} = 0, \;
 \end{align} where $T$ is also a multiple of $m$. 
Therefore we arrive at
\begin{align}
 C_M(\tau) = &\lim_{T \to \infty} \frac{1}{T} \sum_{t = 0}^{T} C_M(t,\tau) \nonumber\\
 = &\sum_k \sum_{p = \pm} e^{- i \omega_p(k) \tau} |\alpha(k,\omega_p(k))|^2 |\psi_+(k, \omega_p(k))|^2 \nonumber\\
 &- e^{- i \omega_p(k) \tau} |\alpha(k,\omega_p(k))|^2 |\psi_-(k, \omega_p(k))|^2~.
\end{align}
 Applying a discrete Fourier transform from the time domain $\tau$ to the frequency domain $\omega$ we arrive at 
  \begin{align}
C_M(\omega) =   \lim_{T \to \infty} \frac{1}{T} \sum_{\tau = 0}^{T} e^{i \omega \tau} C_M(\tau) \nonumber\\
  = \sum_l  |\alpha(k_l,\omega)|^2 |\psi_+(k_l, \omega)|^2 
 - |\alpha(k_l,\omega)|^2 |\psi_-(k_l,\omega )|^2 \nonumber\\
 = \sum_l  |\alpha(k_l,\omega)|^2 M(k_l)~.\label{eq3}
 \end{align} 
In Eq.~(\ref{eq3}) the last sum runs over all such $k_l$ which yield the same frequency $\omega$.  
If the initial state was a superposition of all basis states with coefficients whose absolute values are equal, such that
\begin{align}
 |\alpha(k, \omega_+(k))| = |\alpha(k, \omega_-(k))| = \frac{1}{\sqrt{2N}}~~\text{for all}~k
\end{align}
we finally obtain
\begin{align}
 C_M(\omega) = \frac{1}{2N} M_s(\omega)~.
\end{align}
Note that the assumption of all basis states having the same weight is similar to a generalized notion of infinite temperature. 
In other words $C_M(\tau)$ will be the expectation value of the operator $\left(U^{(m)}\right)^{\tau/m} \cdot \hat{M}$ w.r.t.~a density matrix
$\rho = \sum_k \sum_p \ket{k}\bra{k} \otimes \ket{\psi(k, \omega_p(k))}\bra{\psi(k, \omega_p(k))}$. The $\rho$ is diagonal in eigenbasis with uniform probabilities, 
and hence it describes a thermal state at infinite temperature limit.
The proposed measurement is therefore expected to be capable of detecting a symmetry breaking in the evolution of a quantum Floquet system at
infinite temperature.

\end{document}